\documentclass[apj]{emulateapj}

\begin{document}

\title{The Dynamics of Stellar Coronae Harboring Hot-jupiters I. 
A Time-dependent MHD Simulation of the Interplanetary Environment in the HD~189733 Planetary System}

\author{O. Cohen\altaffilmark{1}, V.L. Kashyap\altaffilmark{1}, J.J. Drake\altaffilmark{1},  
I.V. Sokolov\altaffilmark{2}, C. Garraffo\altaffilmark{1}, T.I. Gombosi\altaffilmark{2}}

\altaffiltext{1}{Harvard-Smithsonian Center for Astrophysics, 60 Garden St. Cambridge, MA 02138}
\altaffiltext{2}{Center for Space Environment Modeling, University of Michigan, 2455 Hayward St., 
Ann Arbor, MI 48109}

\begin{abstract}

We carry out the first time-dependent numerical MagnetoHydroDynamic
modeling of an extrasolar planetary system to study the
interaction of the stellar magnetic field and wind with the planetary
magnetosphere and outflow.  We base our model on the parameters of
the HD 189733 system, which harbors a close-in giant planet. Our simulation reveals 
a highly structured stellar corona characterized by sectors with 
different plasma properties.  The
star-planet interaction varies in magnitude and complexity, depending
on the planetary phase, planetary magnetic field strength, and the relative orientation of the stellar and planetary fields.  It also reveals a long, comet-like tail 
which is a result of the wrapping of the planetary magnetospheric tail by its fast orbital motion.
A reconnection event occurs at a specific orbital phase, causing
mass loss from the planetary magnetosphere that can generate a
hot spot on the stellar surface. The simulation also shows that the system has sufficient energy to produce hot-spots observed in Ca~II lines in giant planet hosting stars. 
However, the short duration of the reconnection event suggests that such SPI cannot be observed persistently.

\end{abstract}



\section{INTRODUCTION}
\label{sec:Intro}
   
The newly discovered exoplanets are not only newly discovered worlds literally, but are also new worlds in terms 
of the physical system they introduce us to. Many exoplanets have now been observed 
since their first discovery \citep{mayor95, exoplanet95, exoplanets03}. In particular, many Jupiter-like giant planets 
have been observed at a distance of less than 0.1~AU from their parent star, and some 
even within $10R_\star$, where they are essentially located inside the extended 
stellar corona \citep{exoplanet95, exoplanets03}. It is not unreasonable to suppose that 
these giant planets have a substantial internal magnetic field 
\citep{sanches-lavega04,durand-Manterola09} and a magnetosphere, which could encompass a significant fraction of the extended coronal volume. 
Indeed, the Jovian magnetosphere is the largest entity in our solar system after the solar one \citep{Bagenal04}.

Perhaps the most fundamental difference between the close-in planet scenario and magnetospheres in our own solar system is that 
close-in planets and their magnetospheres can be located within the stellar Alfv\'en radius.  In analogy to the 
hydrodynamic sonic point (where the flow speed equals the sound speed), the Alfv\'en radius (or the Alfv\'enic point) 
is the distance at which the accelerating stellar wind equals the Alfv\'en speed, $u_A=B/\sqrt{4\pi\rho}$. Here 
$B$ is the magnetic field strength and $\rho$ is the plasma mass density. Beyond the Alfv\'en point, the flow is 
super-Alfv\'enic and magnetic information (or energy) cannot propagate back towards the star and affect the corona.  Close-in planets could then have a substantial effect on the structure of the stellar corona, 
and the electromagnetic interaction between the planetary magnetic field and the corona might generate some 
observational signatures.

In recent years, some signatures of star-planet interaction (SPI) have been 
observed. In a series of papers, \cite{shkolnik03, shkolnik05a,shkolnik05b,shkolnik08} 
have described the observed modulations in the Ca II K emission line, which is an 
indicator of chromospheric activity. They found a line intensity increase corresponding to chromospheric 
``hot spots'' with a period that is correlated with the planetary orbital motion on HD~189733, HD~179949, 
$\tau$ Boo, and $\nu$ And. 
In some cases the location of these hot spots was aligned with the star-planet 
radial vector, while in other cases, the location of the intensity peak was shifted from 
this vector by 70-170 degrees. Some of these observations also revealed an on/off nature \citep{shkolnik08}. 

At higher energies, 
\cite{kashyap08} performed a statistical survey of systems with 
close-in giant planets, which revealed that the X-ray flux of these systems is 
about 30-400\% higher than the typical fluxes from similar stars with planets 
located further out in the stellar system.
\cite{Poppenhaeger10} did not find a similar effect for a
sample of nearby stars, though they did uncover weak evidence for a combined effect
of planetary mass and orbital distance.  \cite{saar08} observed the HD~179949 system in X-rays and found that 
emission associated with a hot spot likely engendered by SPI contributed 
$\approx$30\% to the emission at a mean plasma temperature of $\approx$ 1 keV.  A similar 
trend in the UV band has been recently presented by \citep{shkolnik10}.

Stellar coronae are dominated by the stellar magnetic field, which also dictates the structure of the stellar wind \citep{parker58}.  
The presence of a close-in planet with a significant internal magnetic field could perturb this stellar field and might affect the 
large-scale structure of the stellar corona. The planet can affect the corona in different ways. 
First, a purely hydrodynamic affect is the planet acting as an obstacle, deflecting 
the flow of plasma (i.e., the stellar wind) around it.  Second, an electrostatic effect is
the large-scale potential field topology being modified by the superposition of the stellar and 
planetary magnetic fields. Third, a MagnetoHydroDynamic (MHD) effect is expected due to the orbital motion 
of the 'external' field (the planetary field) inside the background conducting plasma of the stellar 
corona and the stellar wind. In any case, the process involves a modification of the coronal magnetized medium via 
energy transfer between the planetary magnetic field and the corona. 

Few theoretical SPI models have been developed to explain the observed hot 
spots, as well as the overall increase in observed X-ray flux \citep{cuntz00,cranmer07}. 
In particular, \cite{lanza08,lanza09} proposed a mechanism to explain the observed phase shift in the location 
of the spots.  This work also demonstrated that energy transfer from the magnetic field to the plasma via magnetic reconnection between 
the coronal and planetary field (as the planet is moving through the corona) can provide the observed overall energy increase 
(about $10^{21}~W$).  This amount of energy requires rather strong stellar and planetary magnetic fields or alternatively, 
this energy can be provided assuming the planet triggers magnetic reconnection in a stellar corona that is 
already energized due to accumulation of magnetic helicity provided by photospheric motions.

Performing numerical simulations of SPI is challenging. First, one needs to 
provide a dynamic model for the ambient stellar corona and stellar wind. Second, it is necessary 
to introduce the planet, which is an additional boundary condition in the model. 
Third, for capturing the dynamics due to the planetary motion, one needs to consider the planet as a {\it time-dependent} boundary condition. 
\cite{Ip04}, \cite{preusse06}, and \cite{preusse06} have studied the structure of Alfv\'en wings in the  close-in planet around 
HD~179949 and made the analogy of the magnetic interaction between Io and the 
Jovian magnetosphere. \cite{lipatov05} and \cite{Johansson09} have studied the structure of the planetary 
magnetosphere using a hybrid code driven by an approximation to the stellar wind as a 
boundary condition. In both cases, the simulations included only the planetary 
magnetosphere. 

In a recent paper, \cite{cohen09} (hereafter Paper I) have presented two simplified MHD simulations of 
SPI. In the first case (``Case A''), both the stellar and planetary magnetic fields 
were represented by magnetic dipoles, and the planet was fixed in the inertial frame 
of reference. A relative motion due to stellar rotation 
of the coronal plasma with respect to the planetary 
magnetic field was obtained.  In the second case (``Case B''), 
the planetary magnetic field was dipolar again, while the stellar magnetic field was 
obtained using high-resolution {\sl solar} magnetic field data. This provided a realistic stellar 
field, in contrast to the idealized dipole field used in Case A. In Case B, the planet 
was fixed the whole time and the simulation was calculated in the frame of reference 
that is rotating with the star in a tidally-locked manner \citep[like the $\tau$ Boo system,][]{Butler97}. 
The synthetic X-ray light curves of Case A revealed that there is a drop 
in the X-ray flux when the planet is behind the star. In Case B, the 
intensity drop was shifted by $\approx\;60^\circ$ from the star-planet plane. The 
simulation also showed an increase in the total X-ray flux compare to a reference case 
without a planet. 

In paper I, we concluded that the increase in X-ray flux in 
systems with close-in planets is due to  the planet 
preventing the corona from expanding, so that field lines that would be open remain closed. 
As a result, the plasma cannot escape and the overall coronal density is higher than in the 
case without a planet, or with a planet located further from the star. Since the X-ray 
flux is essentially a line of sight integration of the square of the (electron) density, the 
observed X-ray 
flux is higher in such systems. In addition, in paper I we proposed that the 
observed hot-spots are associated with the magnetic connectivity between the star 
and the planet, and that the shift in the location of the hot-spots from the 
star-planet vector is probably due to the complexity of the stellar magnetic field.

In a follow-up study to Paper I, \citet[][hereafter Paper~II]{Cohen10c} investigated the influence of changing the semi-major axis 
of the co-rotating planet on the stellar mass and angular momentum loss. The disruption of the corona and wind occurs when the 
separation is sufficiently small that the planetary and stellar Alfv\'en surfaces
start to interact.  We found  that the
spin-down of stars harboring close-in planets is reduced compared to that of stars with distant planets or no planets at all.

While the results of the simulations presented in Papers I and II are significant, 
both cases were stationary and did not capture all of the dynamical effects that are 
predicted in the theoretical descriptions of SPI (see above). In particular, magnetic reconnection 
between the stellar field and the coronal field was not well captured, even in Paper~I Case A where some relative 
motion between the magnetic bodies was introduced. In this paper, we present a 
dynamical MHD simulation, where we include {\it time-dependent} circular orbital motion of 
the planet around the star. The model is based 
on the observed properties of the HD 189733 system.  However, we emphasize that the 
main goal of this study is to investigate the interaction of a close-in 
giant planet with the stellar corona of its host star as a fundamental plasma 
physics problem, rather than to model HD~189733 specifically.

The structure of the paper is as follows. We present the numerical model in 
Section~\ref{sec:Model}. The results are presented in Section~\ref{sec:Results}, and
our main findings are discussed in Section~\ref{sec:Discussion}.  The results
are summarized in Section~\ref{sec:Conclusions}.
  

\section{NUMERICAL SIMULATION}
\label{sec:Model}

We perform the numerical simulation using the BATS-R-US global MHD model, originally developed 
for the solar corona and the solar wind \citep[][Solar Corona module 
of the Space Weather Numerical Framework]{powell99,toth05,cohen07}. This 
model is driven by synoptic maps of the photospheric radial magnetic field (magnetograms),  
which are routinely observed on the Sun. Similar observations are available for some stellar systems 
using the Zeeman-Doppler-Imaging (ZDI) technique 
\citep[][see \S\ref{sec:StellarField}]{DonatiCollierCameron97,Donati99}. 
Constrained by this boundary condition for the magnetic field, as well as the stellar surface 
plasma density, $\rho_0$, surface temperature, $T_0$, mass, $M_\star$, radius, $R_\star$, 
and rotation period, $P_\star$, the model solves the set of conservation laws for the mass, 
momentum, magnetic induction, and energy (the MHD equations). The stellar 
parameters of HD 189733 used in this simulation are listed in Table~\ref{table:t1}. 
The end result of the simulation is a self-consistent solution 
for the solar/stellar corona and solar/stellar wind, where the volumetric heating needed for the wind 
acceleration is specified according to the magnetic field distribution and an empirical relation 
between the solar wind terminal speed and the magnetic flux tube expansion \citep{wang90,argepizzo00}. 
The implementation of the solar model for the stellar coronal case can be found in paper I, 
\cite{cohen10a}, and \cite{cohen10b}.  We refer the reader to these references for a detailed 
description of the numerical approach taken in the simulation described here. 

\subsection{THE STELLAR MAGNETIC FIELD}
\label{sec:StellarField}

\begin{figure*}[h!]
\centering
\includegraphics[width=6.in]{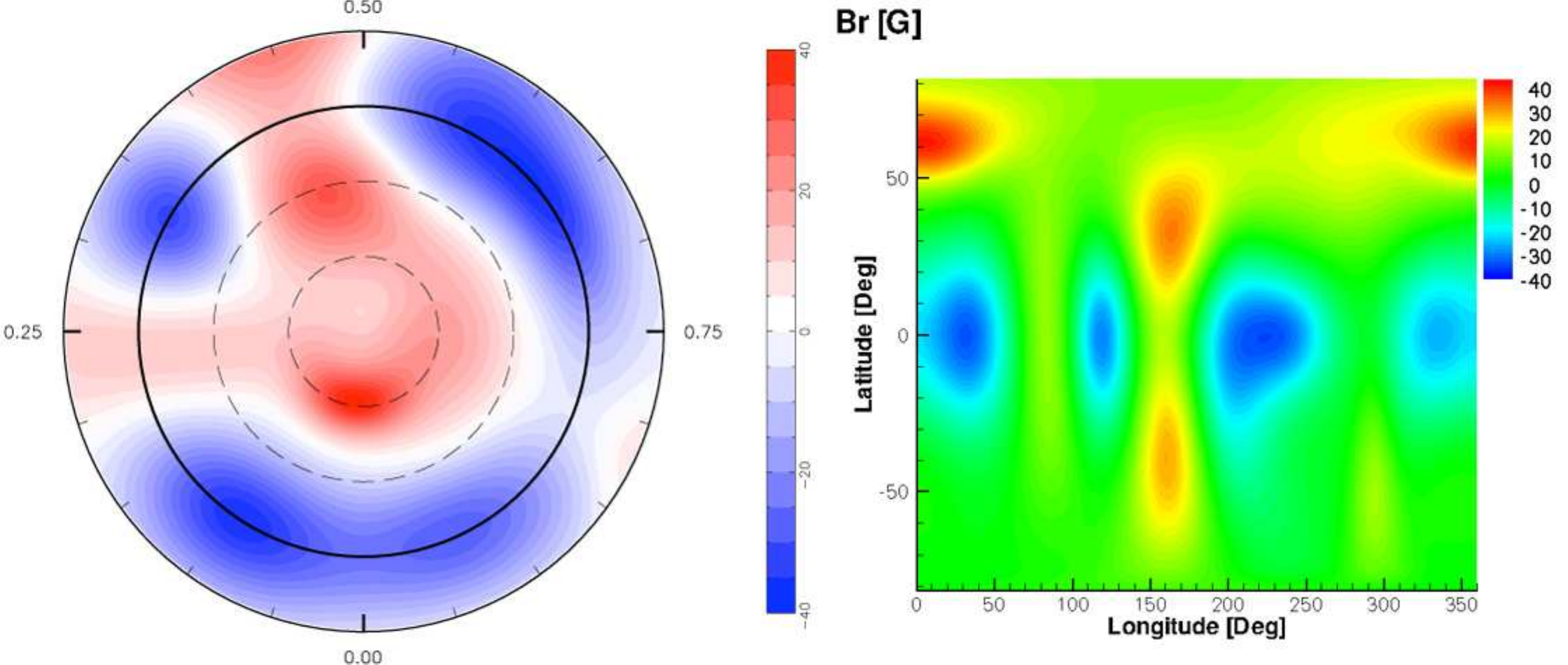}
\caption{A map of the radial stellar magnetic field reproduced from \cite{fares10} in a polar projection (left), and in a longitude-latitude projection (right).}
\label{fig:f1}
\end{figure*}

In paper I, we approximated the stellar magnetic field 
as a dipole. However, recent observations by \cite{fares10} provide us with a better description 
of the stellar field in the form of stellar magnetograms of HD 189733. These measurements were taken during 
June and August 2006, June 2007, and July 2008. In this work, our main goal is to study the 
dynamic interaction between the planetary magnetosphere and a complex, realistic stellar magnetic 
field as the former passes through different sectors of the corona exhibiting different magnetic 
field and stellar wind properties. Therefore, 
we have adopted magnetograms based on the data presented by \cite{fares10}. These low-resolution maps 
do not contain any fine structure, so they are sufficient for our goal 
We base our simulations  on the synoptic map for 2006, since this 
is the epoch when observations by \cite{shkolnik08} were made.  Figure~\ref{fig:f1} shows the 
 the map used in our simulation; this is directly comparable to that illustrated in the top-left of Figure~5 of \citet{fares10}. 
Due to the inclination of the star, the map is incomplete in its southern hemisphere. However, the missing data 
occupies minor parts of the total area, so we choose to superimpose the map with a weak 
background $5~G$ (polar) dipole field, similar to that of the Sun, oriented with positive at the North pole, to complete 
the map.    

\begin{figure*}[h!]
\centering
\includegraphics[width=6.5in]{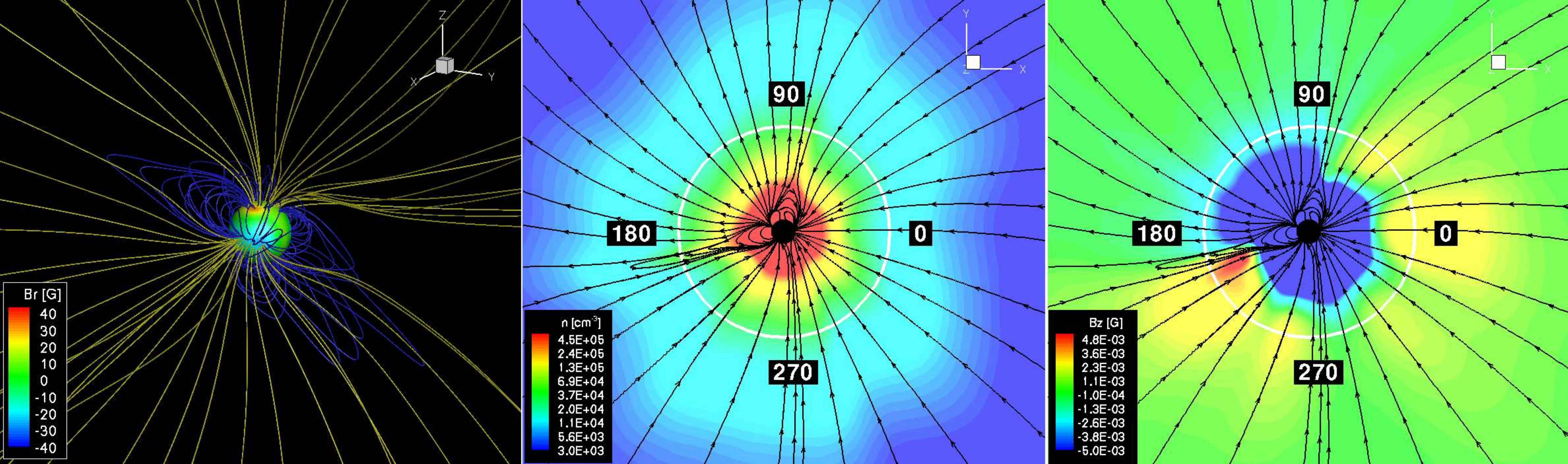}
\caption{Left: steady-state MHD solution driven by the surface field map without the planet. The stellar surface is shown as a sphere 
colored with contours of the radial magnetic field. Blue lines represent closed field lines and yellow lines represent open field lines. 
number density (middle) and $B_z$ (right) color contours are displayed on the equatorial plane along with magnetic filed lines shown as black 
streamlines. Selected phase angles of the planetary orbit are shown as well.}
\label{fig:f2}
\end{figure*}

Figure~\ref{fig:f2} shows the steady-state MHD solution for the adopted magnetogram {\it without} 
the planet. It shows the three-dimensional structure of the stellar coronal field, as well 
as the equatorial plane colored with contours of number density, $n$, and the $B_z$ component of the field. 
Phase angles of 0, 90, 180, and 270, which refer to the reference positions of the planetary orbit, are also displayed.
In the remainder of this paper, we use 
this orientation to describe a particular phase angle. 
Since the planetary magnetic field is a perfect dipole directed towards the North stellar pole, 
the map of $B_z$ shown in Figure~\ref{fig:f2} is used as the stellar reference when describing the inclination 
between the stellar and the planetary field. 

To act as a source of reference in helping to understand the role of the complex surface magnetic topology in the simulations {\em including} the planet, we also computed a simulation for a simplified dipolar stellar magnetic field configuration, with an equatorial field strength of 5~G.
\newline 

\subsection{MODELING THE PLANET}
\label{sec:planet} 

\begin{figure*}[h!]
\centering
\includegraphics[width=6.in]{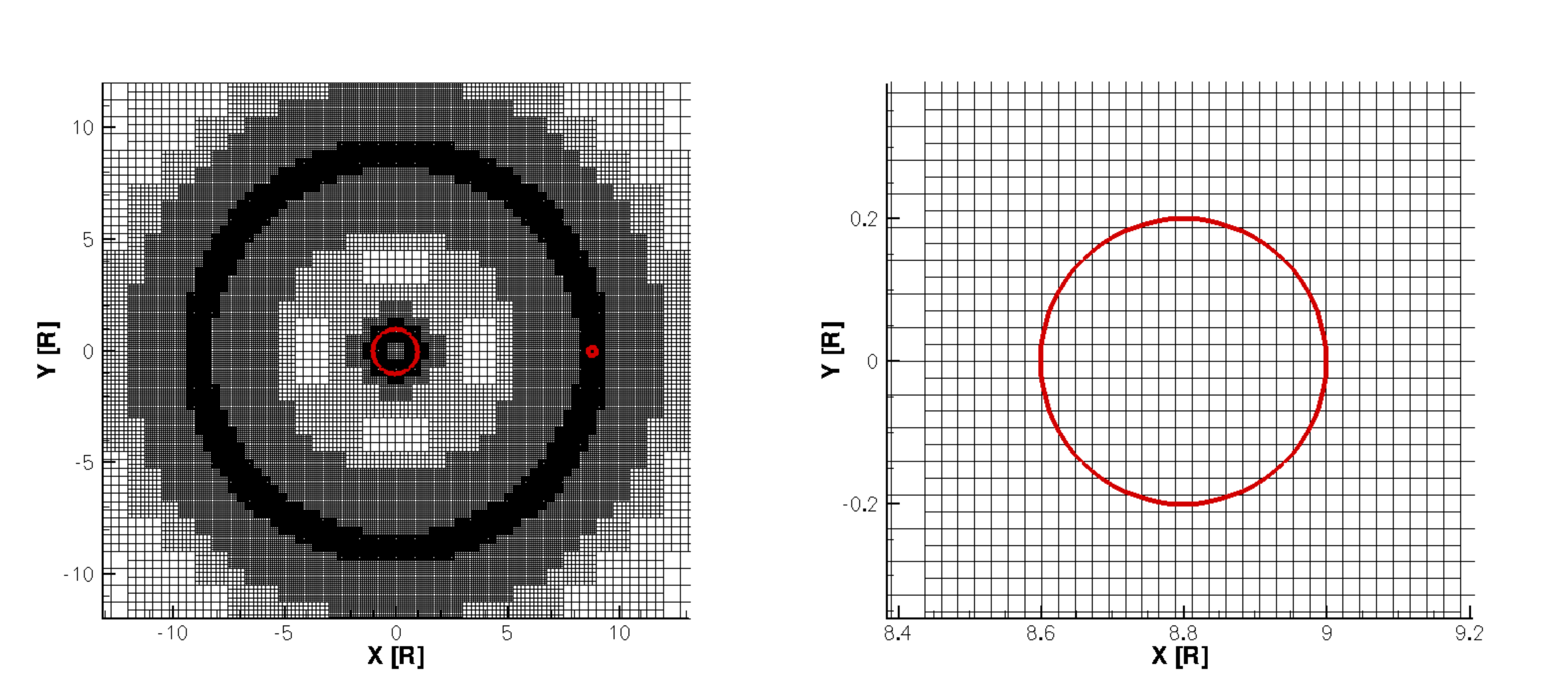}
\caption{Left: the grid structure adopted for the simulation including the planet displayed on the equatorial plane in black, with the star and the planet shown as red circles. Right: 
a zoom-in on the vicinity of the planet.}
\label{fig:f3}
\end{figure*}

In our simulation, the planet is represented by an additional set of boundary conditions for a second body 
(the first body being the star). In order to include the orbital motion in the simulation, the coordinates of 
this second body are updated according to the time-dependent planetary orbital motion. Here we assume 
a circular orbit in the equatorial plane, but in principle any inclined orbit can be specified. 
At each time step, the grid structure is updated to determine which grid cells are boundary cells and 
which cells are real (non-ghost) cells.  The frequency of the updating of coordinates is constrained 
by the numerical stability condition of the simulation (the relation between the grid size, $\Delta x$, 
the time step, $\Delta t$, 
and the Alfv\'en speed).  In addition, the grid resolution 
around the second body should be high enough so that the body is well-resolved. For this reason, 
we choose a planetary radius of $R_p=2R_J$, with $R_J$ being the radius of Jupiter, instead of the observed value of 
$1.1R_J$.  The grid size 
around the second body is $\Delta x=0.025R_\star$ with 18 grid cells across the body. Using the 
flexible grid capabilities of the model, we also used this high-resolution gridding 
around the stellar surface 
(where $\Delta x=0.04R_\star$), and along the planetary trajectory. Figure~\ref{fig:f3} shows the grid structure 
on the equatorial plane with the two bodies marked in red.

The strength of the planetary field is unknown. On the one hand, based on 
stellar evolution models the internal magnetic fields of hot Jupiters are 
expected to be lower than that of Jupiter due to tidal locking and a consequently longer rotation period \citep{sanches-lavega04}. 
On the other hand, a strong field may be required to shield these gas giants from erosion by the strong stellar wind. 
Here we study two set of parameters for the planet. A ``Strong'' case, where we use a dipole equatorial field of $2~G$, 
and a ``Weak'' case, with a dipole equatorial field of $0.1~G$. While we use a fixed boundary condition for the planetary temperature, 
$T_p=10^4\;K $ \citep{murray-clay09}, the boundary condition for the surface density enables us to control the 
planetary thermally driven outflow. This outflow is obtained in the simulation due to the fact that the second body 
serves as a permanent mass source in an environment with lower density. Here we choose a boundary value 
of $n_p=10^9\;cm^{-3}$ for the ``Strong'' case and a boundary value of $n_p=10^7\;cm^{-3}$ for the ``Weak'' case. 
In Section~\ref{sec:Discussion} we discuss the importance of this free parameter in the solution.


\section{RESULTS}
\label{sec:Results}

A salient characteristic of the simulations carried out here is the dependence of the results on orbital phase. 
We emphasize two main aspects here.  Firstly, we find that when the planet is not co-rotating, as was assumed in Paper~I, 
it can no longer provide a steady means of preventing the wind-driven plasma from expanding outward.  While the planet still 
acts to inhibit angular moment loss, a consequence of the interaction is that this varies through the orbital cycle. 
Secondly, magnetic reconnection occurs at certain phases in the combined stellar and planetary magnetic fields that allows 
planetary gas to escape the form its magnetospheric confinement.

In Paper I, we considered the case of a planet that was tidally-locked---fixed in location relative
the stellar surface and co-rotating with the stellar magnetic  field.  This led to the
stellar wind being permanently obstructed in the direction of the planet and the plasma here 
being bottled up. When the planet moves relative to the stellar field, the
wind in the direction of the planet does not remain bottled up, but can slip past once the planet has moved on. Consequently,
the amount of material that is confined by the presence of the planet is reduced, and the structure of the ambient corona and 
stellar wind is quickly recovered.

In Paper II, we found that the stellar mass and angular momentum loss rates are reduced due to the planetary effect on the stellar corona. 
Once the planetary and stellar Alfv\'en surfaces start to interact, the coronal and wind structure is
disrupted by the obstacle---the planet---so that the mass flux carried by the stellar wind decreases. This leads to a
long-term feedback from the planet on the star in the form of a reduction in the stellar angular momentum loss rate to the wind, so that the
spin-down of stars harboring close-in planets is not as high as it would be without the planet, 
consistent with observations \citep{Pont09,lanza10}. This study again considered planets co-rotating with the stellar surface. In the
simulations here, that restraint has been removed.

\subsection{ORBITAL PHASE-DEPENDENT MASS LOSS}

\begin{figure*}[h!]
\centering
\includegraphics[width=5.75in]{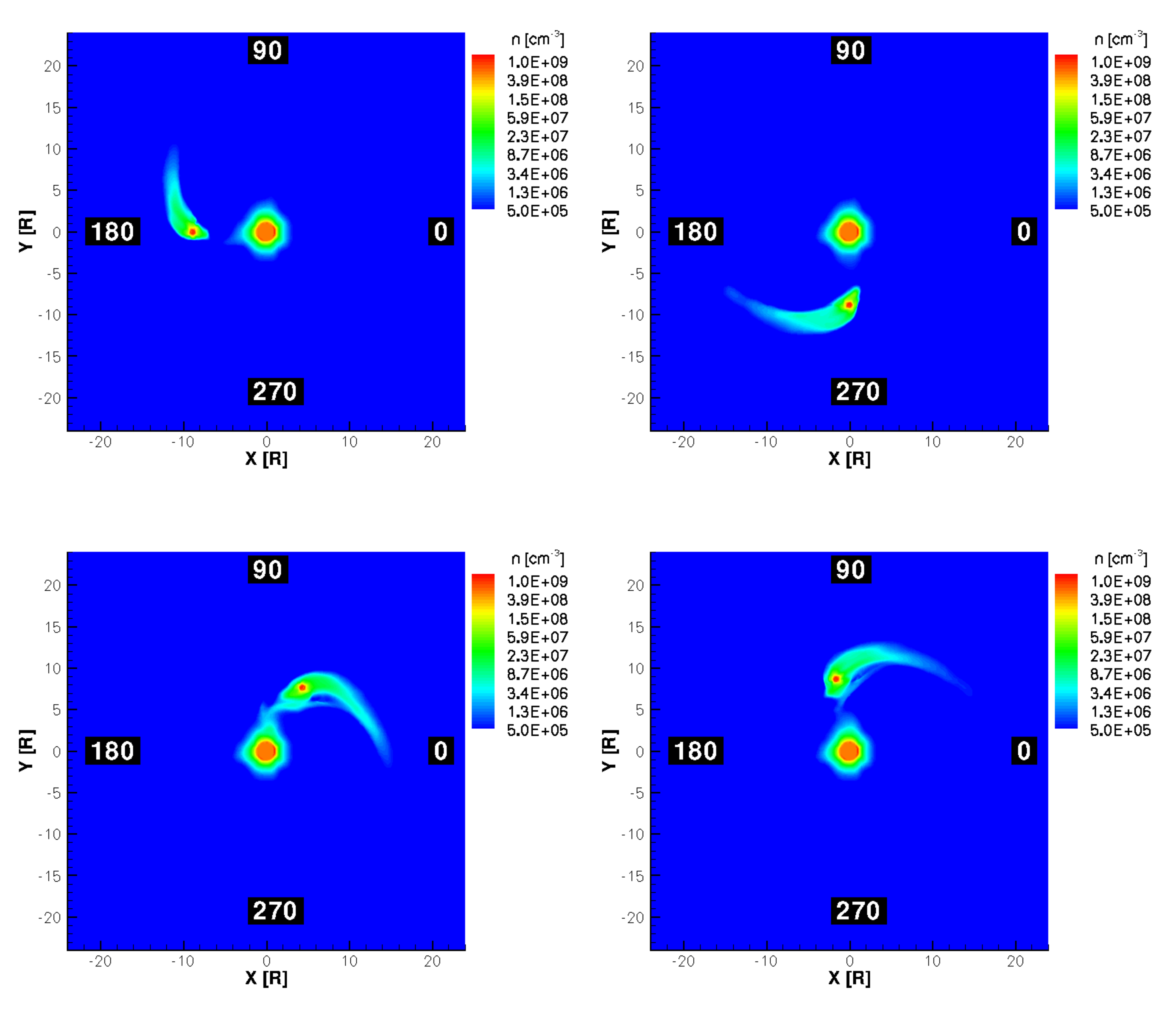}
\caption{Color contours of the number density shown on the equatorial plane for selected phase angles for the ``Strong'' case. 
Axes are expressed in units of stellar radii.}
\label{fig:f4}
\end{figure*}

The gas number density from the simulation at four different orbital phases is illustrated in Figure~\ref{fig:f4} for the ``Strong'' case.  
The salient feature of the 
density distribution is the presence of a comet-like tail of plasma trailing the planet. This feature has been observed by 
\cite{Vidal-Madjar94} and has been predicted from hydrodynamic simulations by \cite{Schneiter07}. 
We will discuss this feature in detail in \S~\ref{sec:planmag} below. In Figure~\ref{fig:f5}  we show the
total {\it outward} mass flux carried by the stellar wind as a function of planetary orbit phase angle that underlies the behavior seen in Figure~\ref{fig:f4}. 
We choose to calculate the mass flux through three spheres in order to illuminate the dependence of the mass loss as a function of the distance from the star.  The first sphere is of radius $r=6R_\star$, which crosses the region between the star and the planet, 
the second is for $r=10R_\star$, which encompasses the planet but crosses the planetary magnetosphere, and the third is for $r=23R_\star$, which occupies most of the simulation domain. 
At $r=6R_\star$, most of the mass flux is carried out by the stellar wind, with a minor contribution from the planet. The mass flux at $r=10R_\star$ 
is dominated by the planetary contribution (in addition to stellar wind mass source), and this is the reason for its relatively high values. 
However, most of this plasma is actually trapped inside the magnetosphere so that it does not contribute to the total mass flux of the system---the apparent outward flow from the planet here is balanced by a similar flow inward. 
The total outward mass flux at $r=23R_\star$ is instead again dominated by the stellar wind, except for phase angles of 0 - 120$^\circ$, where a magnetic 
reconnection event (\S~\ref{sec:magrecon}) allows plasma to escape from the planetary magnetosphere 
and to be added to the total radial mass flux.  Most of this plasma falls onto the star, and this can be seen in Figure~\ref{fig:f4} as a bridge of higher density joining the star and planet.  However, 
our mass flux calculations indicate that there is also a significant planetary contribution from this reconnection to the total mass flux of the stellar wind. 

\subsection{MAGNETIC RECONNECTION AND PLANETARY PLASMA ESCAPE}
\label{sec:magrecon}

As the planet moves into a region where the stellar field has a 
different structure, reconnection can occur between planetary and stellar fields if they have opposite polarity.
In the ``Strong'' case, this occurs around phase angles of 60 - 120$^\circ$, when the planet
gets into a region of negative $B_z$ stellar field associated with a helmet streamer.  This is seen in Figure~\ref{fig:f2}, that illustrates the solution for the stellar field alone and does not include the planet. 
The planetary plasma is generally confined and prevented from expanding freely by 
the magnetic field of the planet.  The reconnection event releases some of this bottled-up plasma and it subsequently falls onto the star.  This is seen as a high density region between
the planet and the star in Figure~\ref{fig:f4}.  This high density region is absent at phase angles 150 - 300$^\circ$, prior to the reconnection and after the released plasma has dissipated. In the ``Weak'' case, shown in Figure~\ref{fig:f6}, the planetary magnetosphere suffers from stronger 
variations than in the ``Strong'' case as seen by the change in the size of its tail. The highest mass escape from the magnetosphere occurs 
around 130$^\circ$, not 60$^\circ$ (for the ``Strong'' case). This demonstrates how sensitive the ``observable'' 
feature is to the planetary parameters. In this particular simulation, a weaker planetary field with the same stellar field resulted 
in a smaller deviation of the reconnection site from the star-planet axis. This is because a weaker planetary field ``feels'' the opposite 
stellar field later than the ``Strong'' case. 

The planetary outflow and the resulting ``hot spot'' where it falls on the stellar surface for the ``Strong'' case can be
seen in Figure~\ref{fig:f7}, where a snapshot is presented for the time a magnetic reconnection event has been identified. 
A local region of infalling mass is seen clearly in the mass flux plot at the stellar surface. 
For reference, we show a similar snapshot from the ``idealized'' simulation computed for the dipole stellar magnetic field of $5~G$ and 
a planetary boundary number density value of $n_p=10^{10}\;cm^{-3}$.  In this case, there is a continual opening and closing of field lines as the planet revolves about the star such that there is always an open field channel joining the star and planet.  The planetary gas is free to expand along this open field region and results in a continuous infall of matter onto a spot on the stellar surface that lies at a constant phase offset from the planetary orbital phase.  In the lower panels of Figure~\ref{fig:f7} this flow can be seen impacting the star approximately 180$^\circ$ ahead of the planet.

The reconnection event for the realistic stellar magnetogram simulation is shown in detail in Figure~\ref{fig:f8}. The four panels on the left show selected magnetic field 
lines and color contours of 
the number density over the equatorial plane. When the planet is far from the reconnection region (top-left panel, phase angle of 270$^\circ$), 
the field topology is of the ambient stellar field, with some closed field lines of the helmet streamer and 
some open field lines dragged by the stellar wind. At a phase angle of 60$^\circ$ (top-right), the field lines that were originally open are now 
connecting the star and the planet. In addition, some of the closed field lines of the stellar helmet streamer are now open as a result of 
the reconnection in front of the planet. At a phase 
angle of 100$^\circ$ (bottom-left), reconnection occurs behind the planet, closing down some field lines. 
Finally, at a phase angle of 180$^\circ$ (bottom-right), the field topology reverts back to its original, ambient stellar 
field.  The two right panels show the reconnection point in front of the planet (phase angle of 60$^\circ$), 
releasing planetary material onto the star, and the reconnection point behind the planet (phase angle of 100$^\circ$), causing a disconnection 
of a plasmoid from the planetary tail.

\subsection{PLANETARY MAGNETOSPHERE CONFIGURATION}
\label{sec:planmag}

The shape of planetary magnetospheres in our solar system is largely defined by solar wind conditions and, in particular,  by the orientation of 
the Interplanetary Magnetic Field (IMF).  In the case of close-in planets like the system simulated here, the magnetospheric shape 
is defined by the plasma conditions in the particular coronal sector interacting with the planet.  The isolation of the planet from the harsh coronal environment 
depends on the magnetopause location, which is defined as the surface of pressure balance between the stellar 
wind dynamic pressure and the planetary magnetic pressure.  If a planet is located far from the star, the magnetopause is associated with a 
shock in front of it as the stellar wind is slowed down to subsonic speeds. In the case of a close-in planet, the magnetopause can be located 
in regions where the stellar wind never reaches supersonic speed and is still accelerating, so the structure of the surface of pressure balance can be highly complicated.  
For display purposes, we choose instead to show the surface where the radial velocity in the frame of reference centered on the planet vanishes 
in the direction of the star. Alternatively, this is the location where the stellar wind in the direction of the planet vanishes. Figure~\ref{fig:f9} 
shows how this location changes in four selected frames of the simulation (before, during, and after the reconnection event).  It can be 
seen that this boundary moves away from the planet following reconnection due to the plasma escaping from the planetary magnetosphere.   This line never reaches too close to the planet, where the magnetic pressure dominates.  Due to the dipole-like magnetic field 
lines close to the planetary surface, we conclude that the dynamic pressure of the planetary outflow at this region is not significant. 
The boundary value for the planetary density affects mainly the filling time of the planetary magnetosphere after a reconnection event that 
allows magnetospheric plasma to escape. 

Another important magnetospheric feature obtained from our simulation is the long magnetospheric comet -like tail that stretches round almost 90$^\circ$ behind the planet. 
This tail is important since it increases the coronal volume affected by the planet. In our static, tidily-locked simulations, as well as in the interaction of 
solar wind and magnetospheres in the solar system, the magnetospheric tail 
is aligned with the direction of the stellar wind. Here, however, it is perpendicular to the wind direction. Therefore, as long as magnetic reconnection 
does not occur, it serves as an elongated obstacle standing in the way of the stellar wind, and can have a larger impact on the wind structure and mass loss rate. 
The convection electric field, $\mathbf{E}=-\mathbf{u_{sw}}\times\mathbf{B}$ ($\mathbf{u_{sw}}$ is the stellar wind velocity) in this case should be stronger 
than when the magnetospheric tail is aligned.  The magnetospheres of close-in planets should, then, introduce different configurations and dynamics in terms of polar 
cup potentials, particle drifts, and magnetospheric/stellar wind drag. The interaction of such magntospheres with Coronal Mass Ejections (CMEs) should also be 
quite different from the interaction of CMEs with magnetospheres in the solar system, since the CME here will hit the magnetosphere from its side and not from 
its nose.  These processes are potentially of considerable interest, but are beyond the scope of this paper.  
\newline


\section{DISCUSSION}
\label{sec:Discussion}

Our simulation reveals a highly dynamic interaction between the stellar and planetary field as the planet is orbiting the star. 
This is due to the complex stellar field, as well as the stellar wind topology. The simulation demonstrates that, 
even when we use a low-resolution map for the surface magnetic field, which does not include fine structure and 
small-scale active regions, the stellar corona contains sectors with quite different plasma properties.  The interaction 
between the planetary and coronal field changes from one sector to the next, while some of the notable features 
are due to the planetary transition between the different sectors.  This is in contrast to the behavior of the ``idealized'' stellar dipolar field simulation, in which the interaction between fields is, not surprisingly, more continuous in nature, with an open field channel always joining the two bodies.

The dynamical, short-lived nature of the different features of the interaction in the realistic field case suggests that searching for observable SPI signatures 
could be problematic, and that ``static'' theoretical models for the origin of these 
signatures might not be valid.  Based on the simulation, the interaction between the coronal field and the planetary magnetosphere will not in general 
be persistent, at least for systems with short planetary orbital periods.  These are the cases where SPI is expected to 
be the strongest.  In addition, the stellar magnetic field topology determines the 
location where changes in the interaction take place, but this magnetic topology is expected to exhibit secular change within time-scales of 
months or even less. 

The idealized scenario in which the stellar and planetary fields are described as perfect, opposite dipoles can explain persistent 
particle acceleration as a result of magnetic reconnection between the two fields \citep{Ip04,cuntz00,cranmer07,lanza08,lanza09}. 
A tidally locked system can produce a plasma density enhancement in closed coronal loops that would be open without the planet (paper I).  However, 
based on our simulation, none of these proposed scenarios can be persistent for the case of a close-in planet that is not tidally-locked with 
the stellar rotation.  Moreover, recent papers \citep{Zarka07,Vidotto10} have suggested that reconnection events between stellar and planetary 
fields can generate radio emissions that can be observed and constrain the magnitude of the planetary field. 
These calculations assume that the accelerated electrons, which are the source for the radio signal, 
appear during the period of time when the planetary and stellar fields are exactly opposite (in particular, 
during a geomagnetic event, when these fields can be opposite for several days). Our simulation shows that for the general case the situation 
of opposite fields is likely to be much shorter in the case of a realistic, more complex stellar field, and that reconnection events 
last for an hour or two at the most.  If radio emissions from such events can be detected at all, they will require long-term monitoring observations 
in order to distinguish such events from flares and other background radio signals. 

Despite its short duration, the reconnection event leads to a release of an amount of plasma from the planetary magnetosphere (the exact 
amount, of course, depends on the particular set of planetary boundary conditions).  Based on the mass flux displayed in Figure~\ref{fig:f7}, 
we define an average value of $F=10^{-11}~g~s^{-1}~cm^{-2}$ and a flux channel 
cross section of $A=\pi(0.1R_\star)^2$ to determine the total planetary mass flux, $\dot{M}_p=F\cdot A=1.5\cdot 10^{-14}~M_J~yr^{-1}$. 
The average mass flux for the ``Weak'' case is about two times lower than the ``Strong'' case. 
If we assume that such reconnection events occur every orbit (2.2 days), the planetary magnetosphere loses $2.5\cdot 10^{-12}~M_J$ 
in a single year  ($\sim 2.5\cdot 10^{-15}~M_\odot ~yr^{-1}$)---not a significant amount integrated over the lifetime of the planet. 
The reconnection depends on the structure of the stellar magnetic field, which is not constant.  Therefore, 
there are likely to be periods during which more reconnection events occur, and periods 
with fewer events.   We further stress that this mass flux is of ``cold'' plasma escaping from the planetary magnetosphere and is not an attempt to model the actual atmospheric escape from 
the planet.
Nevertheless, these reconnection 
events should include particle acceleration and precipitation, as well as radiation enhancement that might impact the atmospheric escape in general. 
In addition, stellar coronal mass ejections could drive similar reconnection events on top of the reconnection between the ambient planetary and static 
stellar fields.  Such effects, however, are out of the scope of this paper and we defer further discussion to future work. 

It is tempting to associate the infalling planetary plasma with the hot spot discovered in the Ca II K band by 
\cite{shkolnik03, shkolnik05a,shkolnik05b,shkolnik08}, 
but we caution that our simulations do not describe the precise mechanism by which energy
is transferred from the planetary orbital motion to the stellar corona.  The volumetric integral of the sum of 
the magnetic and the kinetic energy in the simulation domain is 
\begin{equation}
E_{tot}=\int \left( \frac{B^2}{8\pi}+\frac{\rho u^2}{2} \right)dV.
\end{equation}
Throughout the simulation, the value of this integral for the ``Strong'' case ranges between (5.4 - 5.6)$\cdot10^{36}~ergs$, with a change of the order of 1\%. 
The change in the energy for the ``Weak'' case is about the same. 
The total volumetric energy for the MHD solution {\it without} the planet is $5.2\cdot10^{36}~ergs$ so that the contribution of the planetary orbital 
motion and the planetary magnetic field is about $2\cdot10^{35}~ergs$. Since the MHD model is limited 
in its description of magnetic reconnection, it can only produce the change in magnetic field topology, which is the source for modulations in 
the total energy. In reality however, magnetic reconnection between the stellar and planetary fields will involve dissipation of this energy, 
as well as non-thermal particle acceleration. Therefore, the 1\% difference in the total energy, $E_{avail}=5\cdot10^{34}~ergs$, is the available energy 
for these mechanisms. The timescale, $\tau$, for the reconnection event in our simulation is about a quarter of an orbit, or $5\cdot10^4~s$, 
so the total available power is:
\begin{equation}
P\approx E_{avail}/\tau=10^{30}~ergs~s^{-1}.  
\end{equation}
Assuming an efficiency of 1\% in converting this power, our simulation finds at least $10^{28}~ergs~s^{-1}$ is available to accelerate particles along 
field lines connecting the planet and the star.  These accelerated particles then hit the chromosphere and cause the observed hot-spots. 
This amount of energy is comparable with the total X-ray power of HD~189733 \citep{Pillitteri10}, and it is reasonable to believe that 
it can produce the observed modulations in the Ca II band, which are also estimated to involve energies of the order of 
$10^{28}-10^{28}~ergs~s^{-1}$ \citep{shkolnik05a,Hall07}.
Variations in the total mass flux at different radial distances in the simulation domain (see Figure~\ref{fig:f5}) reveal that the planetary magnetosphere contributes to the 
total mass flux during the reconnection event. The total mass flux through the Alfv\'en surface based on the simulation without the planet is 
$8.3\cdot 10^{-14}~M_\odot~yr^{-1}$, which is comparable to, but slightly smaller than, the values from the simulation that includes the planet. This differs from the result 
obtained in Paper II for a synchronously revolving planet: in that case, the magnetosphere served only as an obstacle that reduces the amount of stellar 
wind that crosses the Alfv\'en surface. Here however, the planetary magnetospheric shape can be changed by the dynamics of the system, it can contribute to 
the total mass flux, and it disrupts the wind only for a short duration of time. It seems that for the particular set of parameters of the 
HD~189733 system, the effect of the planet on the stellar wind mass loss rate is not that important.  



\section{SUMMARY and CONCLUSIONS}
\label{sec:Conclusions}

We have performed a realistic MHD simulation of SPI using the parameters of the HD 189733 planetary system. 
The simulation is based on the observed stellar surface magnetic field and includes the stellar wind, time-dependent planetary orbital motion and 
planetary outflow. We choose to study two cases. One with strong planetary magnetic field and high base density and one with weak planetary field 
and low base density. Based on the simulation results, our main findings are:
\begin{enumerate}
\item The use of a realistic stellar field results in a highly structured stellar corona characterized by sectors with 
different plasma properties.
\item As the planet orbits the star, it interacts with the different sectors.  Each sector crossing, 
as well as the transitions between the sectors, leads to changes in the SPI.
\item The planet is followed by a long, comet-like tail of plasma that represents its magnetospheric tail wrapped around due to the 
rapid planetary orbital motion. This tail is almost perpendicular 
to the stellar wind direction and can lead to magnetospheric dynamics which differ to those in the case in which the magnetospheric 
tail is aligned with the wind direction.
\item A large reconnection event occurs when the planetary and stellar field are opposite in polarity, where the reconnection site varies 
depends on the planetary field strength. This reconnection event causes a release of cold plasma from the planetary magnetosphere with a mass flux of 
$\dot{M}_p=2.5\cdot 10^{-12}\;M_J\;yr^{-1}$. However, this escape of material can be defined as a magnetospheric 
escape rather than atmospheric escape. The latter cannot be define by the simulation presented here since it does not include a 
self-consistent atmospheric escape mechanism and is only controlled by a boundary value for the planetary density;
\item The short duration of the reconnection event suggests that such SPI cannot be observed persistently, and in 
particular when such signatures are being searched for in radio bands.
\item The simulation indicates that a sufficient amount of energy can be dissipated by the star-planet interaction to explain the hot-spots observed in the Ca~II lines of stars hosting close-in planets. 
\item The total mass flux carried by the stellar wind is modulated by the planetary orbit and the particular value at any time depends on the magnetic topology. 
\end{enumerate}

In this work, we used the parameters of HD 189733, with planetary semi-major axis of $a=8.8R_\star$. It is reasonable to believe that 
with a smaller value of $a$, some of the SPI effects in the simulation would be enhanced.  Further studies of SPI involving 
a physical atmospheric escape from the planet, as well as the impact of stellar coronal mass ejections on the planetary magnetosphere, would be highly motivated.


\acknowledgments
We would like to thank Scott Wolk and Steve Saar for their useful comments. 
OC is supported by SHINE through NSF ATM-0823592 grant, and by NASA-LWSTRT Grant NNG05GM44G.
JJD and VLK were funded by NASA contract NAS8-39073 to the {\it Chandra X-ray Center}.
Simulation results were obtained using the Space Weather Modeling
Framework, developed by the Center for Space Environment Modeling, at the University of Michigan with funding
support from NASA ESS, NASA ESTO-CT, NSF KDI, and DoD MURI.




\begin{table}[h!]
\caption{Stellar and Planetary Parameters of HD 189733}
\begin{tabular}{cc}
\hline
Stellar Parameter & Value\\
\hline
$R_\star$ & $0.76\;R_\odot$ \\
$M_\star$ & $0.82\;M_\odot$ \\
$P_\star$ & $11.95\;d$  \\
\hline
Planetary Parameter & Value\\
\hline
$R_p$ & $1.1\;R_J$$^a$  \\
$P_{orb}$ & $2.2\;d$  \\
$B_{eq}$  & $2.0\;G$ \\
$a$ & $8.8\;R_\star$ \\
\hline
\end{tabular}
$^a$A value of 2 has been adopted in this simulation)
\label{table:t1}
\end{table}


\clearpage

\begin{figure*}[h!]
\centering
\includegraphics[width=6.5in]{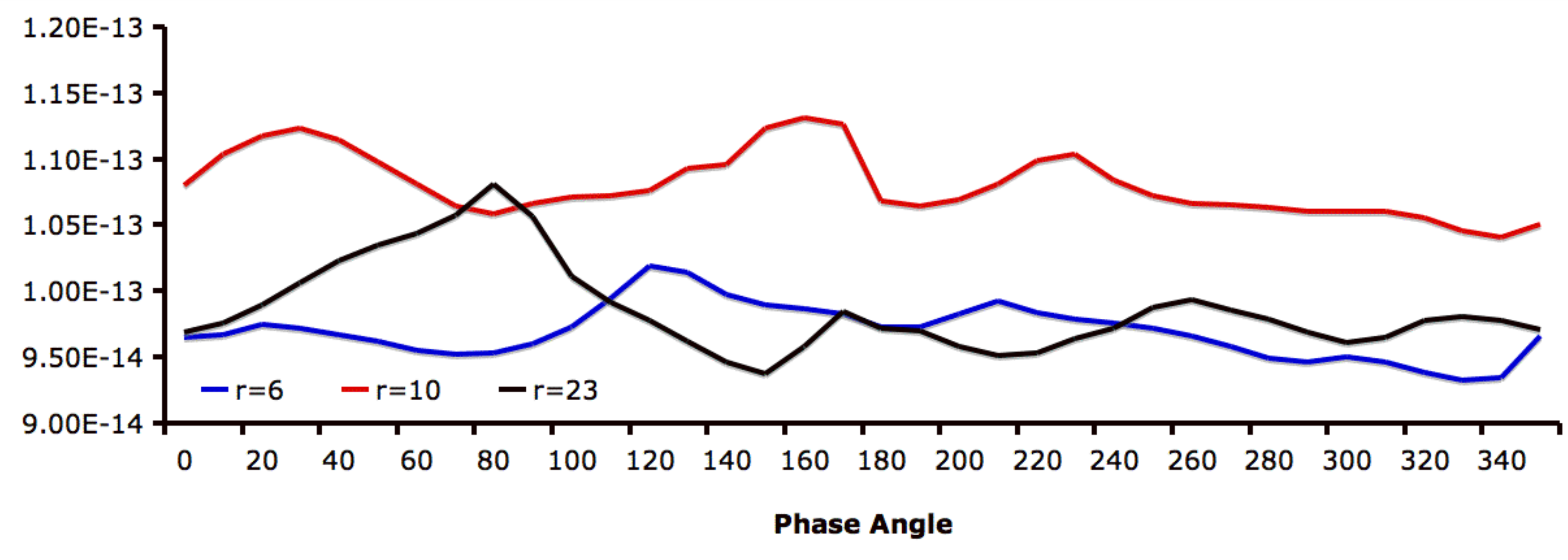}
\caption{Mass flux as a function of planetary phase angle integrated over spheres of $r=6R_\star$ (blue), $r=10R_\star$ (red), 
$r=23R_\star$ (black).}
\label{fig:f5}
\end{figure*}
\clearpage

\begin{figure*}[h!]
\centering
\includegraphics[width=5.75in]{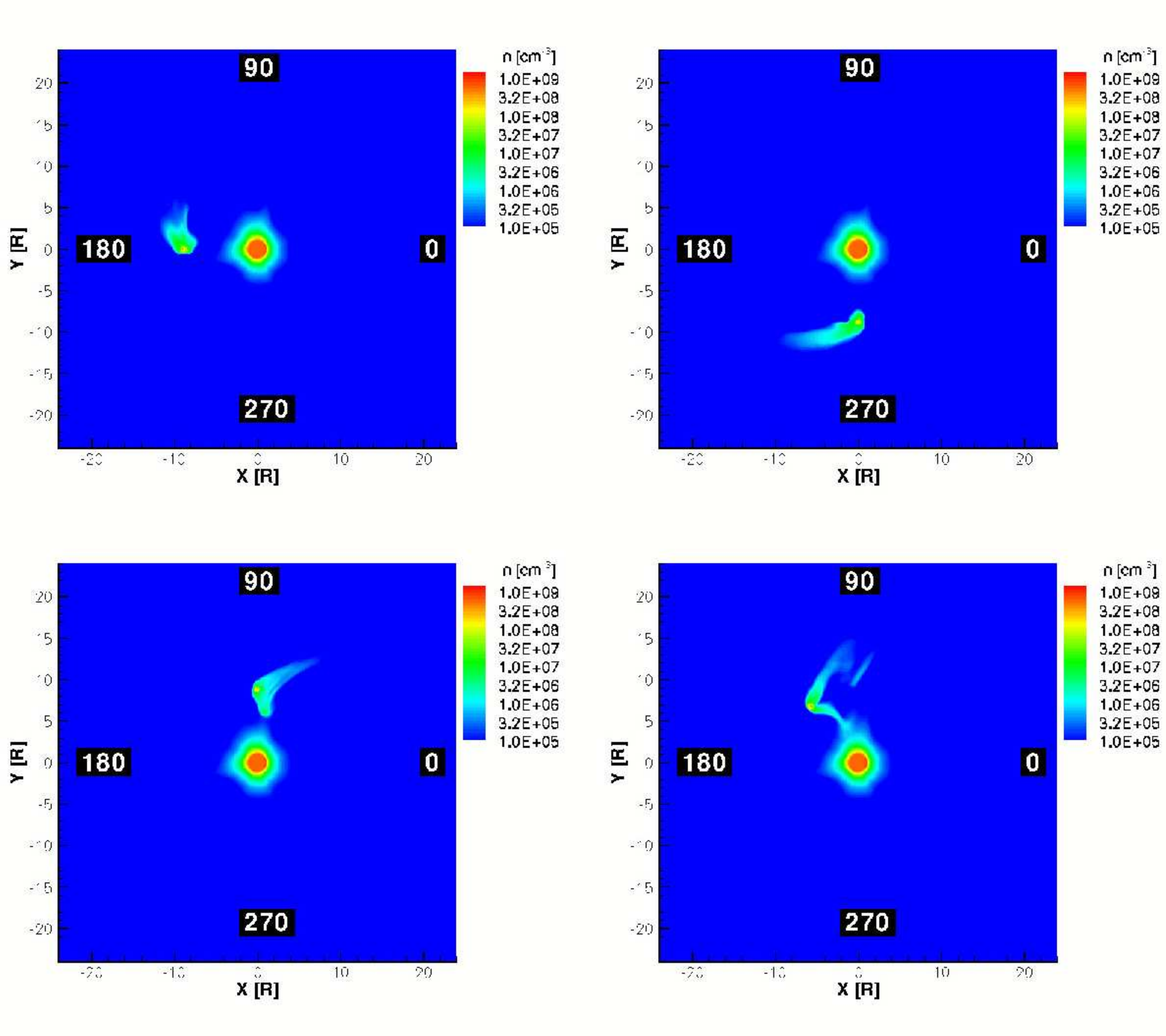}
\caption{Same display as Figure~\ref{fig:f4} but for the ``Weak'' case. Note that the minimum density value here is slightly lower than 
Figure~\ref{fig:f4}.}
\label{fig:f6}
\end{figure*}
\clearpage

\begin{figure*}[h!]
\centering
\includegraphics[width=6.5in]{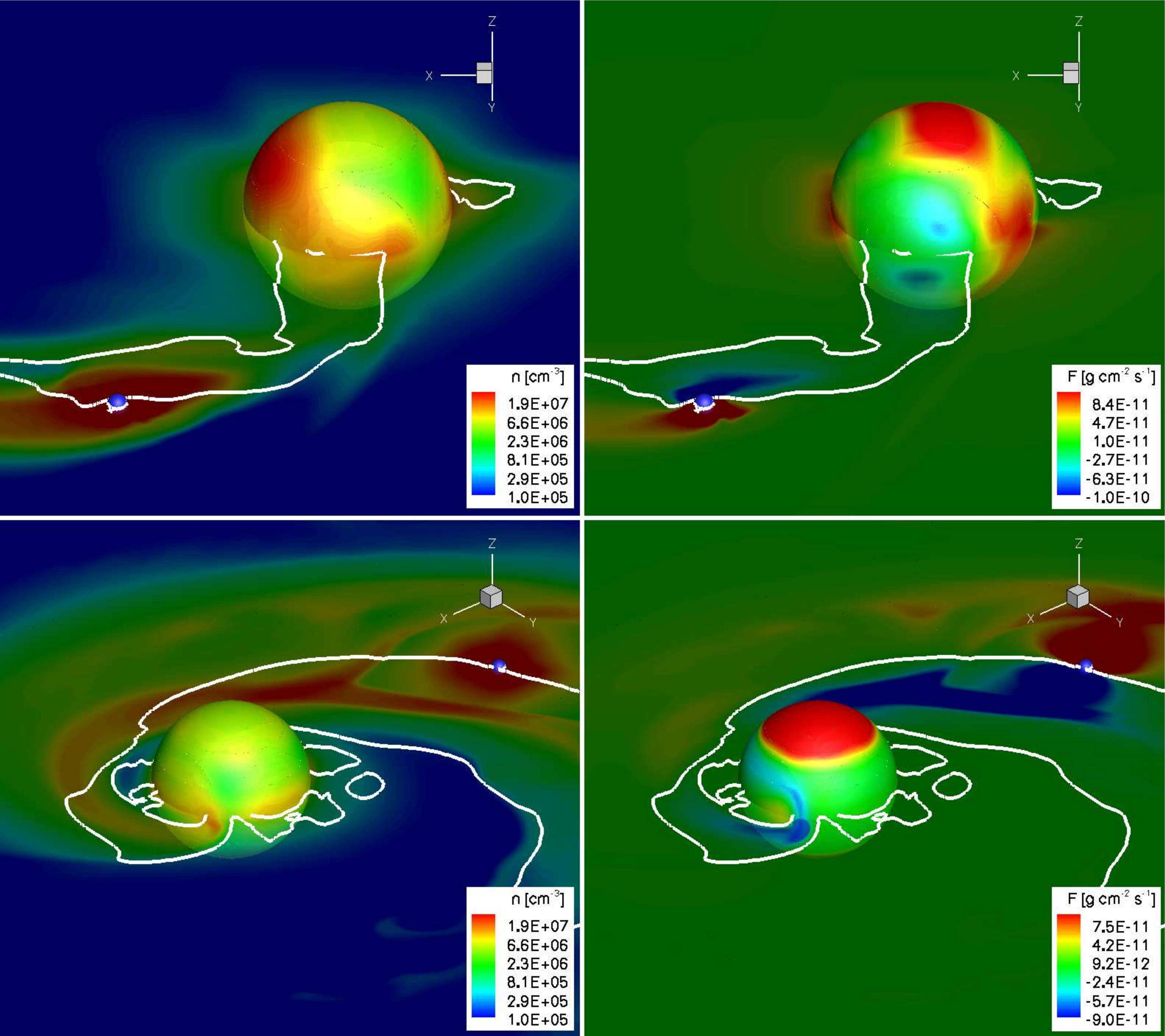}
\caption{Color contours of number density (left) and mass flux (right) displayed on the equatorial plane and on a sphere of $r=2R_\star$, 
with the planet shown as a blue sphere. White line separates positive and negative values of the mass flux. 
The plots are for the dynamic simulation for the time when a magnetic reconnection event has been identified, 
at phase angle of 60 degrees (top), and for the idealized simulation at a similar phase with an enhanced planetary surface density 
and dipolar stellar field (bottom).}
\label{fig:f7}
\end{figure*}
\clearpage

\begin{figure*}[h!]
\centering
\includegraphics[width=6.in]{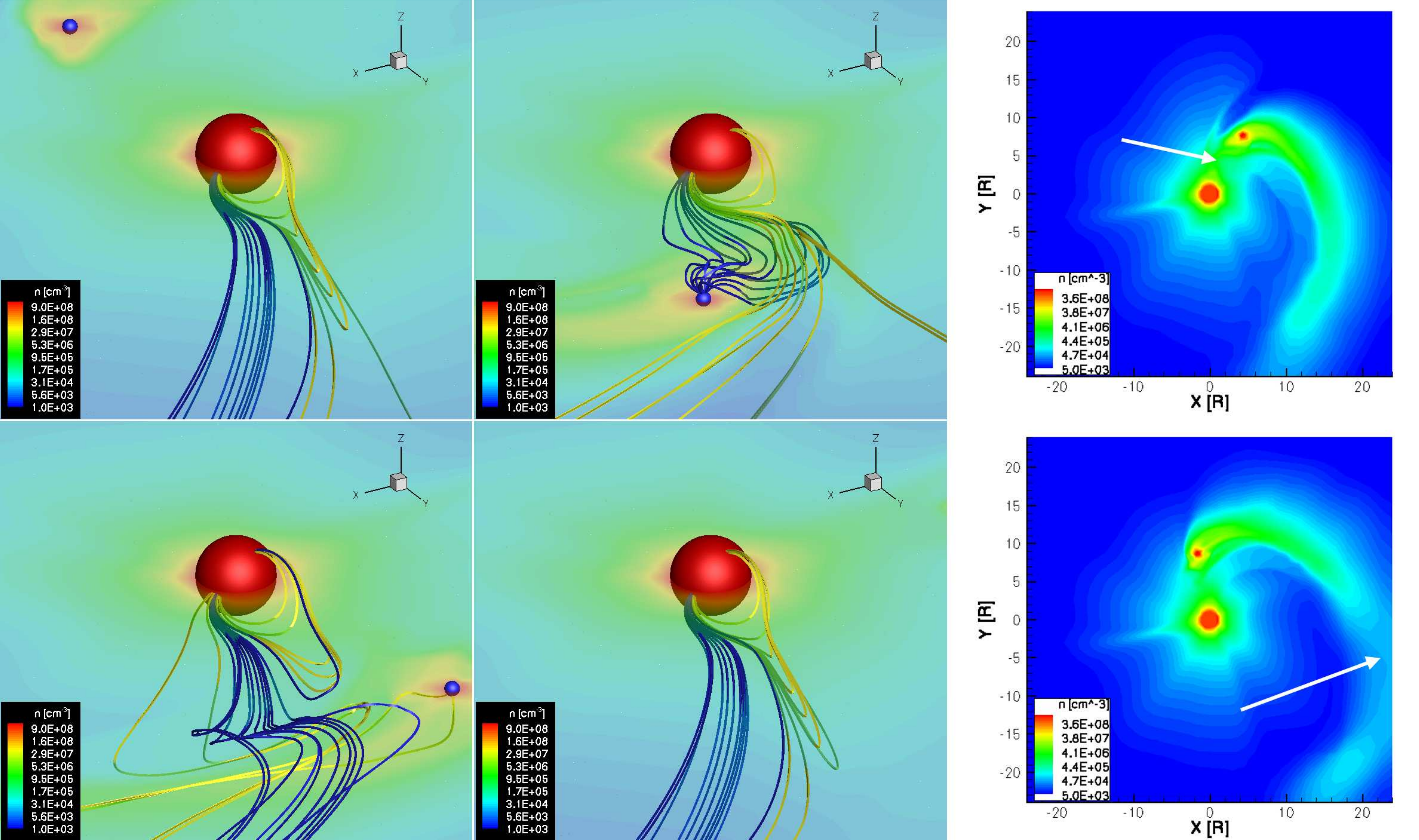}
\caption{Left four panels: the equatorial plane colored with number density contours, together with red and blue spheres representing the star 
and the planet, respectively. The panels are for phase angles of 270 (top-left), 60 (top-right), 100 (bottom-left), and 180 (bottom-right) degrees. 
Selected magnetic field lines are shown in yellow and blue. Two right panels: the equatorial plane colored with number density contours for 
phase angle of 60 degrees(top) and 100 degrees (bottom).  The white arrows indicate the regions where reconnection is taking place.}
\label{fig:f8}
\end{figure*}

\begin{figure*}[h!]
\centering
\includegraphics[width=5.in]{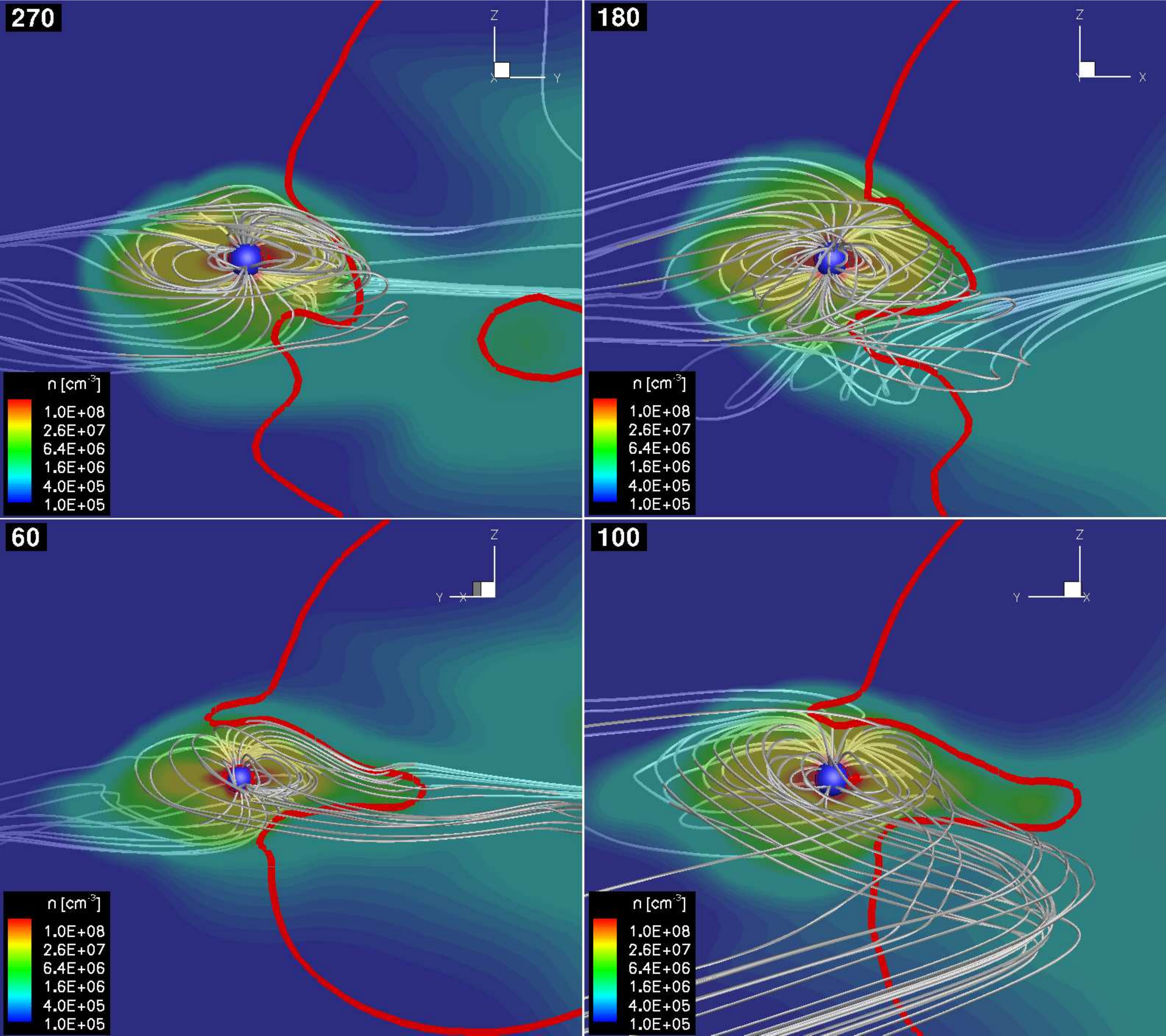}
\caption{Magnetospheric boundary at different phase angles defined by vanishing radial stellar wind.   In each case the star lies to the right, out of the frame.  The boundary is shown as a solid red line displayed 
on the star-planet meridional plain along with color contours of number density. Selected magnetic field lines are shown in white.  Illustrated are phase angles of 270 (top-left panel), 180 (top-right panel), 60 (bottom-left panel), and 100 
(bottom-right panel) degrees.}
\label{fig:f9}
\end{figure*}

\end{document}